# Clustering kinetics during natural ageing of Al-Cu based alloys with (Mg, Li) additions

R. Ivanov, A. Deschamps and F. De Geuser

Univ. Grenoble Alpes, CNRS, Grenoble INP, SIMAP, 38000 Grenoble, France.

## Abstract

Room temperature solute clustering in aluminium alloys, or natural ageing, despite its industrial relevance, is still subject to debate, mostly due to its experimentally challenging nature. To better understand the complex multi-constituents' interactions at play, we have studied ternary and quaternary subsystems based on the Al-Cu alloys, namely Al-Cu-Mg, Al-Cu-Li and Al-Cu-Li-Mg. We used a recently introduced correlative technique using small-angle neutrons and X-ray scattering (SANS and SAXS) to extract the chemically resolved kinetics of room temperature clustering in these alloys, which we completed with differential scanning calorimetry (DSC) and micro-hardness measurements. The comparison of the clustering behaviours of each subsystem allowed us to highlight the paramount role of Mg as a trigger for diffusion and clustering. Indeed, while a strong natural ageing was observed in the Al-Cu-Mg alloy, virtually none was shown for Al-Cu-Li. A very slight addition of Mg (0.4%) to this system, however, drastically changed the situation to a rapid formation of essentially Cu-rich hardening clusters, Mg only joining them later in the reaction. This diffusion enabling effect of Mg is discussed in terms of diffusion mechanism and complex interactions with the quenched-in vacancies.

## 1. Introduction

Natural ageing of Aluminium alloys encompasses the changes at room temperature to the microstructure and related properties after they have been quenched from solution treatment temperature. First observed by Wilm in 1906 [1] and later characterised by Guinier and Preston [2,3], the increase of hardness experienced during natural ageing at room temperature is related to the changes in solute distribution at the atomic scale. Solute clustering describes the initial stage of decomposition of a supersaturated solid solution into one with solute-rich domains retaining the host crystal structure. It results in measurable changes of the materials properties, in particular hardness, but also electrical resistivity, lattice parameter etc.

Solute clusters, being of the same crystal structure as the host Al matrix, showing no interface and only mild compositional contrast have posed great difficulties to be unambiguously characterized and render necessary the combined use of several analytical techniques [4]. Most common experimental techniques based on global sample methods such as differential scanning calorimetry (DSC)[5–10], electrical resistivity [8,9] and of course hardness [11–14] provide evidence for the formation and evolution of clusters. Direct imaging of Cu-Mg "clusters" has been reported by Ralston et al. by high-resolution aberration-corrected TEM [15]; however this method has proven very difficult due to the small quantity and low Z-contrast of the solute species associated in individual clusters with large number density. Techniques based on positron annihilation spectroscopy [4,16–23], nuclear magnetic resonance [24,25] or X-ray absorption spectroscopy [18,26] provide more information about the local environment of solutes and can shine some light on solute-vacancy relationships. Atom Probe Tomography (APT), giving access to the 3D position and chemical identity of atoms, has been extensively used in conjunction with cluster-finding algorithms or radial distribution functions (RDFs) to provide information on the correlation of solutes and size distributions of clustered domains [27–42]. Small angle scattering (SAS) techniques, providing a global view of the decomposition of the solid solution, have been until now only sparsely used to study clustering [4,25,43,44], despite them being used extensively to describe the degree of inhomogeneity of solute during spinodal decomposition, such as in the Fe-Cr system [45–50].

Since the diffusion of all major solute elements in Aluminium is substitutional, clustering kinetics depends on the presence of vacancies, their supersaturation resulting from the quench from the solution treatment temperature, and their interaction with the solutes they are conveying as well as with the clusters themselves. These interactions have been particularly studied in Al-Mg-Si alloys [31,4,32,8,17,51,52,9,33,36,37,12,38,20,21,39,23,13,14,40,10] owing for the prominent – generally negative – effect of clustering on the subsequent artificial ageing kinetics sought during paint baking cycles for automotive sheet. An extensive picture of clustering in this system has been reached by combining experimental observations, thermodynamic modelling [53], and more recently atomistic modelling [42,54]. These results show that clusters are generally present as early as the as-quenched state [21] and then evolve in number density, size but also chemistry, with several families of clusters involving several types of solutes coexisting in the microstructure. These characteristics depend of course on the alloy chemistry, with noticeable effects having been demonstrated by adding controlled amounts of minor solutes effective at trapping vacancies and thus delaying the natural ageing kinetics [12,13].

The other landmark system for clustering studies is Al-Cu-Mg, in which Mg and Cu form co-clusters that are very effective at increasing the alloy's strength. Only a few seconds at moderate artificial ageing temperatures (150-200°C) are sufficient to obtain up to 70% of the maximum precipitation strengthening potential of such alloys, hence the name "rapid hardening" for this effect of clustering [5,27–29]. Clustering kinetics in this system has also been evaluated by combinations of APT, DSC, PALS, and modelling [16,6,7,55,26,56–59,18,11,35,42,41]. Despite a weakly attractive interaction between Cu and vacancies and a weak repulsion between Mg and vacancies [60], the addition of Mg to Al-Cu considerably accelerates the natural ageing process [26,11]. "Mg-vacancy complexes" (more on this expression in the discussion) have been claimed to already form during quenching [16,19,22]. However, clusters are mainly rich in Cu in the as-quenched state and incorporate Mg gradually during natural ageing [7,55]. Meanwhile, vacancies are surrounded by solute atoms and seem to become gradually trapped within clusters during ageing [18]. However, due to the low concentration of quenched-in vacancies (of the order of $10^{-5}$ according to [56]), most clusters are rapidly vacancy-free.

A successful model for cluster growth has been developed considering that the limiting process was the detachment rate of vacancies trapped in these clusters [58]. The interaction of vacancies with solutes during the clustering process seem to be better described by non-trivial vacancy-clusters interactions rather than the sum of pairwise vacancy-solute bindings [42].

Al-Cu-Li-Mg alloys have been the object of intensive research in the last 10 years for aerospace applications [61,62]. Despite considerable work devoted to describing the precipitation sequence in these alloys as a function of thermo-mechanical history and alloy composition [63–72,43,73–77], the study of natural ageing has been so far quite limited. In the high Li-containing alloys developed initially, most of the low-temperature phenomena have been identified to precipitation of the $\delta'$-$Al_3Li$ phase [78,79]. However, in many of the recently developed Al-Cu-Li alloys, lower levels of Li do not result in appreciable amounts of this phase [43,80] and the natural ageing has been shown to involve Cu-rich clusters [81,43]. Recently, we have set up a correlative methodology for the quantitative evaluation of the quantity of solute atoms involved in clusters and the corresponding Mg/Cu ratio using X-ray and neutron small-angle scattering (SAXS & SANS), supplemented by atom probe tomography [44]. Using this methodology, it has been possible to demonstrate that a small addition of Mg to Al-Cu-Li profoundly changes the clustering behaviour of Cu atoms [44]. The study of a compositionally graded Al-Cu-Li-(Mg) alloy with varying Mg concentration has shown that this positive effect of Mg on the clustering rate of Cu monotonically increases with Mg concentration [82].

In order to understand the respective roles of Li, Cu and Mg on clustering in this quaternary alloy, it is useful to evaluate separately the role of the different solutes and thus compare the different ternary alloys Al-Cu-Mg, Al-Cu-Li to the quaternary alloy Al-Cu-Li-Mg. Using the recently developed correlative SAXS / SANS methodology on three such alloys, the aim of the present paper is to provide a quantitative characterization of the chemistry (Mg/Cu ratio) and quantity of clusters continuously during natural ageing. These results will be supplemented by a more classical DSC evaluation, hence providing a better understanding of the heat events that have been already observed by previous studies.

## 2. Materials and experimental methodology

Three subsets of the Al-Cu system were chosen to study the clustering behaviour during natural ageing. The alloys tested have nominal compositions shown in Table . The starting material was provided as 27 mm thick plates by Constellium-Voreppe C-TEC, France. The plates were homogenized at 500 °C for 24 h, quenched and subsequently hot rolled to 3 mm at 350 °C. Flat samples from the 3 mm plates were ground and polished for micro-hardness observation using a 200 gf Vickers indenter. They were solution treated at 500 °C for 30 min, water quenched and tested. The measurements obtained within 5 min after quench correspond to the as-quenched condition (AQ). Hardness evolution was monitored during the following 2 weeks. Samples aged for 72 h or longer are referred to as naturally aged (NA).

Differential scanning calorimetry samples were prepared from the rolled plates by cutting 3 mm diameter discs at 0.5 mm thickness by a slow speed saw. The samples were batch solution treated at 500 °C for 30 min, water quenched with the as-quenched (AQ) samples being tested within 2 min. Additional samples were left at room temperature after quench for 1, 2, 4, 6, 16 and 72 hours to represent steps along the ageing process. DSC experiments were carried out using a TA instruments Q200 DSC equipped with an RCS 90 cooling system allowing experiments to start at -50 °C. The experiments used pure Al crucibles and a 10 °C/min heating rate for all runs. As clustering reactions usually take place at room temperature, they are expected to generate formation and dissolution peaks at temperatures below 300 °C at the rate tested in the DSC. To focus on the low-temperature clustering region and maximize the quality of the baseline correction so as to measure very weak clustering peaks, a three-step DSC procedure was used similarly to that used in [80]. It consists of 1) heating ramp from -50 to 350 °C, 2) cooling segment down to -50 °C at 50 °C/min and 3) final re-heat up to 350 °C of a sample (versus an empty crucible). This procedure generates a sample specific baseline (step 3) which does not depend on estimates of the heat capacity of the alloy, the reference used, the crucibles used or the sample vs reference position within the DSC. Subtracting the signal of step 3 from step 1 generates accurate baseline corrected data for the clustering reactions taking place.

Small angle neutron scattering (SANS) experiments were carried out at the D11 instrument at the Institute Laue Langevin (ILL) in Grenoble, France under proposal 1-01-142 [83]. Samples from the 3 mm thick plate were solution treated, quenched and measured during natural ageing up to 12 h. Additional samples aged ex-situ were tested for the naturally aged condition (72h of natural ageing). A wavelength of 5 Å was used, above the Bragg cut-off, ensuring that no double diffraction could bring additional noise. The sample to detector distance was kept constant at 1.2 m capturing scattering vectors from 0.07 to 0.6 $Å^{-1}$. All experiment data has been background corrected, normalized to absolute units and azimuthally averaged using the Graphical Reduction and Analysis SANS Program package (GRASP).

Small angle X-ray scattering samples were obtained by grinding and polishing the 3 mm plates down to ~300 µm, solution treating at 500 °C for 30 mins and water quenching. The final thickness reduction to 80 µm was achieved within 15 mins after the quench. These samples were labelled as quenched (AQ) and their evolution was monitored in-situ during 72h up to the naturally aged (NA) condition. X-rays were generated using a Rigaku MicroMax-007 HF rotating anode with Cu $K_α$ X-ray source (λ=1.52Å) and scattering data was collected using a DECTRIS Pilatus 300K detector. The photon counting Pilatus detector, with its virtually zero noise, allowed for excellent statistics even at the low counting rates measured in the present study. The collected data were background corrected and normalized to absolute units with a glassy carbon sample as a secondary calibration standard using in-house scripts. The sample to detector distance was 0.6 m allowing for measurement of scattering vectors ranging between 0.02 and 0.5 $Å^{-1}$. For better comparison with neutron scattering, we chose to convert the X-ray intensities to cm$^{-1}$ rather than the more conventionally used in SAXS (e Å$^{-3}$), which requires a normalization by the scattering cross section of an electron. The combined scattering data from experiments with neutrons and X-rays have been interpreted using the recently proposed methodology described in [44] which yields results as a correlation length ($\xi$) indicative of cluster size, a modulation periodicity ($\lambda$) related to the distance between clusters and a mean square number of excess solutes ($\overline{\Delta n_{Cu}^2}$) involved in clusters.

## 3. Results

The evolution of solute distribution taking place at room temperature during natural ageing was monitored using micro-hardness, DSC, SAXS and SANS. The majority of changes occurred during the first 24 h of ageing and measurements were found to plateau after 72 h resulting in the NA condition. The details of this evolution are described in the following sections.

### 3.1. Hardness

Hardness evolution after quench depicts the undergoing changes in solute distribution from nearly random in the AQ condition to clustered in the NA condition in the alloys studied. The hardness increase at room temperature versus time is shown in Figure 1. The Al-Cu-Li and Al-Cu-Li-Mg alloys both start at 67±2 HV evidencing the negligible effect of adding 0.4 at. % Mg to the Al-Cu-Li in the AQ state. The Al-Cu-Mg has a higher AQ hardness at 74±2 HV due to the higher Mg contribution to solid solution strengthening. These initial hardness values serve as an indication of the degree of solute solution strengthening by alloying elements as no significant amount of clustering is expected this early after the quench.

As natural ageing progresses, alloys containing Mg show a significant increase in hardness compared to the Mg-free alloy. The AQ difference between Al-Cu-Mg and Al-Cu-Li-Mg alloys diminishes for longer ageing times. Hardening begins to saturate after about 48h and for longer ageing times the hardness changes remain within experimental errors around the average NA value. Ageing for 72 h (3 days) after quench has been chosen as a point beyond which all alloys are considered stable and can be termed 'naturally aged'. The NA hardness values are 77±3, 117±3 and 115±3 HV for the Al-Cu-Li, Al-Cu-Mg and Al-Cu-Li-Mg alloys respectively. Exposure to room temperature after quench has on average resulted in a 20, 55 and 82 % hardness increase in these alloys respectively. A vast change in the behaviour of the Al-Cu-Li alloy by the addition of 0.4 at. % Mg is clearly shown and highlights the important role of Mg solute atoms for the natural ageing processes. Furthermore, kinetic and quantitative similarities of the Al-Cu-Mg and Al-Cu-Li-Mg alloys, which contains a significantly different amount of Mg, suggests that only a small quantity of Mg is necessary to achieve cluster strengthening.

### 3.2. Differential scattering calorimetry

The heat flux vs temperature profiles at selected times during natural ageing of the Al-Cu-Mg alloy are shown in Figure 2. The profiles can be divided into two distinct regions with the first containing the exothermic peak between 0 and 150 °C and a second with the endothermic region between 150 and 260 °C. Within the first region, Cu-Mg clusters are forming during the DSC run, which gives rise to the exothermic peak. The potential for cluster formation decreases with ageing time at room temperature as more clusters have formed prior to the DSC scan leading to the magnitude and integral of this region to halve within 1h and continue to decrease during the first 6h. The formation of Cu-Mg clusters observed here is in line with previously reported DSC experiments on Al-Cu-Mg alloys [6,7]. The second region covers the dissolution of clusters previously formed in the sample (whether at room temperature or during the DSC scan). The area of the second region does not vary significantly with ageing time, meaning that the sum of the solute (Cu, Mg) involved in clusters formed at room temperature (during natural ageing) and those involved in cluster formation during the DSC experiment (in the as-quenched sample) is the same. The dissolution of these clusters results in a consistently observable double minimum endothermic region that is characteristic of the Al-Cu-Mg alloy. Variations in cluster chemistry, cluster size distribution, or even a complex dissolution-growth process may lead to such behaviour. Finally, the dissolution of Cu-Mg clusters liberates solute necessary for the subsequent precipitation reactions of the S phase and causes strong exothermic signal for all ageing times at temperatures above 260 °C [72,84].

The temporal evolution of the heat flux profile for the Al-Cu-Li alloy is shown in Figure 3 with the same y-axis (heat flux) range as for Al-Cu-Mg in Figure 2. In this alloy, the profiles remain almost flat with only a small magnitude endothermic region between 50 and 200 °C. No well-defined exothermic reactions are observed during the DSC experiments independently of the ageing time. For longer ageing times at room temperature, e.g. NA condition, the endotherm increases with a broad peak at 125 °C indicative of more dissolution compared to the AQ condition. Therefore, clustering reactions in this alloy are too slow to occur during the DSC scan but can occur during the course of a long natural ageing, yet yielding a much smaller signal as compared to Al-Cu-Mg.

Addition of 0.4 at% Mg to the Al-Cu-Li alloy vastly changes the heat flux profiles occurring at various stages of ageing as shown for the Al-Cu-Li-Mg alloy in Figure 4. Globally, the curves can be divided into two regions with one for the reactions taking place between 50 and 135°C and a second for the reactions between 135°C and 250°C. Within the first region, an exothermic peak is observed for short ageing times and its magnitude is reduced with ageing up to 4h. This peak is in close proximity to that resulting from the formation of Cu-Mg clusters in the Al-Cu-Mg alloy (Figure 2). Ageing for longer than 4h leads to an endothermic peak appearing in the same temperature range. The proximity of the exothermic peak (80 °C) in the AQ condition and the endothermic peak (100 °C) in the NA condition makes it unlikely that the solutes involved in these reactions are the same. The presence of Li in this alloy compared to the Al-Cu-Mg and previously reported the strong influence of Mg on Li ordering in an Al-Li-Mg system [85,86] suggest that the endothermic peak at 100 °C corresponds to disordering of Li short-range order present in the NA condition [80,87,88].

The profiles group up and remain in close proximity during the reactions in the second region, namely the temperature range between 135 and 250 °C. Dissolution of clusters is taking place independently of their formation at room temperature (NA) or during the DSC experiment (AQ). The shape of the endothermic peaks resembles those observed for the Al-Cu-Mg alloy (Figure 2) where Cu-Mg clusters dissolve, albeit at a lower temperature. For temperature above 200 °C, the dissolution reactions approach completion and precipitation starts, mainly of the $T_1$ phase ($Al_2CuLi$) [89] leading to exothermic signal at temperatures above 250 °C.

### 3.3. Small angle scattering

Scattering profiles obtained during the first 10 h after quench and for the naturally aged condition (72h) are shown in Figure 5, which shows together data for X-rays (SAXS) and neutrons (SANS) at the same ageing times. Overall, X-ray scattering results in higher intensity compared to neutrons due to the higher scattering cross-section of X-ray radiation. The X-ray scattering intensity observed is predominantly representative of Cu clustering as this atom yields the highest contrast to the Al matrix (Table 2). Additional and complementary information is derived from neutron scattering where the contrasts of Mg and Cu with respect to Al are comparable. In Li-containing alloys, clustering of Li would result in a high scattering intensity owing to the large contrast between Li and Al. However, Atom Probe Tomography has shown that this solute is not clustered in the states presented [44] so that we will not consider the contribution of Li to clustering from now on. During the ageing period, the scattering intensity for both X-rays and neutrons increases for scattering vectors, q, in the range 0.1 - 0.4 $Å^{-1}$ corresponding to domains with a characteristic size of the order of (10 - 2.5) Å.

In the Al-Cu-Mg alloy, an increase in small-angle scattering intensity is observed in Figure 5a and this behaviour agrees with hardness and DSC results shown above. Development of the clusters leads to a more pronounced intensity change for neutron scattering as compared to X-rays. Since both Mg and Cu contribute to SANS intensity compared to predominantly Cu contribution observable by X-ray scattering, it can be inferred that clustering in the Al-Cu-Mg alloy involves both Cu and Mg atoms.

The scattering profiles for the Al-Cu-Li alloy show only a small degree of evolution during natural ageing for both neutron and X-ray scattering (Figure 5b). Any small degree of scattering for this alloy is due to Cu rich clusters under the assumption that Li atoms remain randomly dispersed. These Cu-rich clusters thus show only a small evolution during natural ageing, in agreement with hardness and DSC results.

Finally, clusters in the Al-Cu-Li-Mg system (Figure 5c) show a drastic evolution despite the only small addition of 0.4 at.% Mg when compared to the Al-Cu-Li alloy. The large scattering intensity appearing in SAXS proves that a large part of this effect is related to Cu clustering. The critical role of Mg for the clustering at room temperature in the Al-Cu-Li-Mg system is evidenced by the dramatic difference between the Al-Cu-Li and Al-Cu-Li-Mg alloys (Figures 5b and 5c).

The model describing the scattering intensity due to concentration fluctuations arising from clusters developed by Ivanov *et al.* [44] has been fitted to the in-situ data for combined neutron and X-ray scattering experiments. The model successfully captures the evolution of intensity during natural ageing as shown by the solid lines in Figure 5. The simultaneous use of complementary scattering from X-rays and neutrons grants access to the chemical nature of clusters, as well as the quantity of solutes involved in the clusters. This quantity, available independently for both Mg and Cu, is expressed as the mean square number of excess solutes involved in the clusters per unit volume, and thus is a measure of the total quantity of solute atoms involved in the clusters irrespective of their composition. In this sense, it is more representative of the degree of completion of the clustering reaction for a given species that of a volume fraction, which in the case of clustering has no clear meaning, in the absence of a well-defined interface, and a possibly evolving chemical composition of forming objects. To further emphasize the chemical evolution of the clusters, the Mg/Cu ratio in clusters can be then calculated and compared to the global alloy ratio.

We first discuss the behaviour of the Al-Cu-Mg alloy. Initially (AQ condition), a small amount of clustering is observed, predominantly Cu rich (Mg/Cu ratio of less than 0.5). Throughout ageing the excess Cu in clusters remains relatively stable, increasing from 5 to 10 $nm^{-3}$, which is consistent with the only moderate increase of SAXS intensity during this ageing period. Conversely, during the first hour of ageing, the excess Mg in clusters increases to match the Cu and surpasses it as shown by the rapid evolution of the Mg/Cu ratio. Initially, clusters contain less Mg compared to the matrix concentration, but when natural ageing progresses, the mean square number of excess Mg increases up to the NA condition reaching 34 ±11 $nm^{-3}$ and representing a ratio of 1.82 to Cu.

Similarly, the initial clusters present in the Al-Cu-Li-Mg alloy are on average domains with excess Cu (Figure 7a) and almost no Mg. The degree of clustering of Mg begins to increase alongside a further increase of the degree of clustering of Cu for ageing times above 2h. Finally, the naturally aged clusters contain on average an excess of 38.9 ±0.6 $nm^{-3}$ of Cu and 9.0 ±0.18 $nm^{-3}$ of Mg resulting in an Mg/Cu ratio of 0.48 that is above the mean ratio of the matrix. In terms of size, the clusters in the Al-Cu-Li-Mg alloy become significantly larger than that of the Al-Cu-Mg alloy

In the case of the Al-Cu-Li alloy, obviously, only the Cu clustering behaviour is measured. Consistently with the results seen before, the clustering kinetics in this alloy is extremely sluggish, and both the size and quantity of clusters is hardly evolving with time.

## 4. Discussion

The results detailed above evidence a series of effects that need now to be rationalized. In terms of strengthening, the ternary Al-Cu-Li alloy shows a very sluggish natural ageing, whereas the ternary Al-Cu-Mg and the quaternary Al-Cu-Li-Mg alloys show a comparable faster strengthening behaviour. In fact, the natural ageing strengthening of the quaternary alloy is larger despite the low Mg content, while an Mg/Cu ratio close to 1 has often be considered necessary to maximize the cluster strengthening effect in Cu and Mg-containing alloys.

The specific feature of our results is to obtain a quantitative evaluation of the clustering kinetics for both Mg and Cu separately. It demonstrates the considerable acceleration of clustering in these alloys brought by Mg addition. Our results evidence in addition that the rates of clustering of Cu and Mg at room temperature are profoundly different. In the two Mg-containing alloys under consideration, the Mg/Cu ratio monotonously increases during ageing. Initially much lower than the alloy composition's ratio, it becomes slightly larger after three days of natural ageing, reaching ~2 in the Al-Cu-Mg alloy and ~0.5 in the Al-Cu-Li-Mg alloy. It is interesting to note that the Mg/Cu cluster ratio reached in these two alloys is quite different from a ratio of ~1 often found in the literature [7,41] and sometimes hypothesized as intrinsic to the formation of Cu-Mg co-clusters [57].

In fact, while the clusters are initially only moderately richer in Cu than in Mg in the Al-Cu-Mg alloy (Mg/Cu ratio of 0.3-0.4), the situation is very different in the Al-Cu-Li-Mg alloy. In this case, during the first 2 hours of natural ageing, the clusters are virtually Mg-free while Cu rapidly clusters, with kinetics comparable to that of the Al-Cu-Mg alloy but up to a much larger amount. It is only when the Cu clustering is more than halfway completed that the Mg starts to be incorporated within the clusters. However, most interestingly, this does not mean that the presence of Mg is unimportant for the initial formation of Cu clusters since, in the absence of Mg, clustering barely happens (Al-Cu-Li). In Al-Cu-Mg alloys, the fact that clustering initially proceeds by Cu atoms and involves later formation of Cu-

Mg rich clusters has been observed by several authors, by APT [7] and by the evolution of lattice parameter [55]. The dramatic effect of Mg alloying on Al-Cu and Al-Cu-Li has been shown by electrical resistivity and hardness [90]. Hirosawa et al. [67] showed that the addition of 0.5% Mg to an Al-Cu-Li alloy accelerated the formation of Cu-rich GPI zones. Klobes et al. [26] showed by X-ray absorption spectroscopy that the rate of Cu decomposition at room temperature was considerably accelerated in the presence of Mg.

The difference in clustering kinetics between the three alloys illustrates the complex interactions between the various solutes, alone or in combination, and the quenched-in vacancies. In this context, the literature often makes use of expression such as "vacancy-solute clusters" or "vacancy-solute complexes". These are possibly misleading in our opinion since 1) there are at least 3 orders of magnitude between any number of solutes and the number of vacancies present at all time and 2) there are at least 5 orders of magnitude between the diffusivity of vacancies and that of solutes. Given these two points, a much better reasoning consists in describing the amount of time spent by the vacancies in the vicinity of certain objects. This is also true for the concept of vacancy trapping which should be understood as some sort of dynamical trapping: each vacancy spends most of the time in the vicinity of the trapping element, but necessarily navigates from one trapping atom to the other. We will use this kind of description in the following.

The easiest case to rationalize is that of the Al-Cu-Li alloy. Well documented in the literature [67,91,92], there is a strong binding energy between Li and vacancies, resulting in the trapping of vacancies by Li atoms after quenching and resulting in sluggish clustering kinetics. The interactions between vacancies and solutes have also been well documented in Al-Cu-Mg alloys. In the as-quenched state, Nagai et al. [16] showed that vacancies were preferentially bound to Mg atoms rather than Cu atoms in Al-Cu-Mg, or rather, spent more time close to Mg atoms than to Cu atoms. In a binary Al-Mg alloy, Zou et al. [19] found so-called vacancy-Mg complexes after quenching, that did not evolve with further natural ageing, which we interpret as the fact that the vacancies were mostly around Mg atoms. This has been confirmed by Liu et al. [22] who further observed a decrease in the vacancy jump frequency. This vacancy-Mg attraction may seem contradictory with the reported slightly repulsive interaction between these two entities [60]. However, it has been shown that Mg-Mg atoms could bind together at a distance of ~0.5 nm [52] so that the binding energy of a vacancy with a group of atoms may be different to that with a single atom. Recent calculations [42] actually proposed an attractive binding energy between vacancies and Cu-Mg clusters, which helps to understand the mechanism proposed by Klobes et al. [18] of a gradual trapping of vacancies within the Cu-Mg clusters after an initial regime where the environment of the vacancies does not change.

While the concept of vacancy-solute binding energy is itself not unambiguous, it does not capture the actual diffusion mechanism which may lead to different situations depending on the actual vacancy jump probabilities around solutes. Because of this, different views can be found in the literature concerning the role of Mg with respect to the concentration of quenched-in vacancies. According to Nagai et al. [16] the migration of Mg-vacancy complexes to dislocations causes a rapid decrease of the vacancy concentration. According to Marceau et al. [11,59], the presence of Mg increases the concentration of vacancy-type defects and reduces the loss of vacancies during ageing. While different mechanisms may be at play, it seems clear that vacancies have an average vicinity which is enriched in Mg. This certainly helps the stabilization of quenched-in vacancies around clustering solutes. This picture does not really explain the formation of the initial Cu clustering, however. This question becomes much more acute in the Al-Cu-Li-Mg alloy.

Indeed, two interesting features in the quaternary alloy combine: first, the dramatically increased clustering kinetics with respect to the Al-Cu-Li alloy with only 0.4 at% Mg addition, and secondly the absence of Mg within the clusters in the first two hours of ageing. The first effect can only be understood if the addition of Mg overcomes the binding of Li to vacancies. Actually, the movement of vacancies during natural ageing does not result in long-range diffusion of Li, as evidence by APT in [44], although some short-range ordering of Li seems to occur, whose reversion results in the endothermic peak at 100°C observed in the DSC scan of the naturally aged Al-Cu-Li-Mg alloy. The effect of Mg on triggering the clustering of Cu has actually been shown to be active even for very small amounts of Mg [82] and to vary monotonously with the Mg concentration. Moreover, Honma et al. [93] have proposed that the addition of Li to an Al-Cu-Mg-Ag alloy decreased the quenched-in vacancy concentration. Thus, from the available knowledge, the effect of the Mg addition seems to replace the Li atoms network as vacancy dynamical trapping sites by an Mg atoms network, or even simply to offer proxy trapping positions in-between Li atoms. Since Mg has a tendency to form energetically stable Cu-Mg bonds, clustering proceeds. Nevertheless, the effective diffusion coefficient of Mg, which depends on the actual jump rates, is lower than that of Cu. This scenario is consistent with the findings of Klobes et al. [18] that vacancies are surrounded both by Cu and Mg atoms already early in the ageing process, despite the clustering involving first mainly Cu atoms. The gradual incorporation of Mg atoms within clusters would then lead to the observed enrichment of the clusters in Mg.

Based on the understanding of the clustering kinetics directly evidenced by the combined use of SAXS and SANS, it is now possible to revisit the interpretation of the DSC curves presented in figures 2 to 4. Concerning the Al-Cu-Mg alloy, the exothermic peak between 40-130°C shows a clear correlation with the clusters that form during natural ageing. Namely, the kinetics of the decrease of the area of this peak (corresponding to the clustering of solutes that have not yet gathered during natural ageing) matches the kinetics of cluster formation. In the case of the quaternary alloy Al-Cu-Li-Mg, the as-quenched state shows an exothermic peak in the same temperature range as compared to the Al-Cu-Mg alloy, suggesting a similar nature of clustering forming in the ternary and quaternary alloys during ramp heating. In the naturally aged condition, the interpretation of the DSC signal of the quaternary alloy is made more complicated by the presence of the endothermic event that we attribute to the formation of short-range ordering of Li (which may also be present to a smaller degree in the naturally aged condition of the Al-Cu-Li alloy). This endothermic event may be already present to some degree at shorter ageing times and thus modify the apparent evolution of the exothermic peak corresponding to the cluster formation. This may explain why the evolution of the exothermic peak seems to occur faster than expected from the cluster formation kinetics measured by small-angle scattering: from DSC, the peak decreases by a factor of 2 in 1 hour, however at that time clustering has hardly started to occur. However, it interesting to note that the endothermic events corresponding to the dissolution of the clusters formed during natural ageing (between 140 and 270°C approximately) are very similar in the ternary and quaternary alloys, despite a very different Mg/Cu ratio. If we combine this information with the evolution of hardness which is comparable in the two alloys, we can conclude that similar evolutions of cluster hardness and thermodynamics of the system can be obtained with very different cluster chemistry and size.

## 5. Conclusion

Clustering kinetics during natural ageing post-quench have been successfully characterised using a combination of differential scanning calorimetry and small angle scattering and complemented with micro-hardness evolution. The results clearly identify Mg as a solute necessary for clustering at room temperature in the Al-Cu-(Li, Mg) system. The small quantity of Cu rich clusters formed during quench from solution temperature does not evolve in the Mg-free solid solution. When the Mg/Cu ratio of the alloy is high, Mg rapidly clusters with the initial Cu rich domains resulting in an increase of hardness due to the formation of strong Cu-Mg bonds. The Mg/Cu ratio of these clusters in the naturally aged condition is close to 2. For the Al-Cu-Li-Mg alloy, where the Mg/Cu ratio is low, clustering is dominated by the formation of Cu rich domains. These domains contain Mg but in significantly lower concentrations with the Mg/Cu ratio remaining below 0.5 for the NA condition, a value which is higher than the nominal Mg/Cu ratio. The combination of these results provides strong evidence for Mg acting in a manner to effectively increase the concentration and/or the mobility of vacancies available for diffusion. These results also suggest that the composition of Cu-Mg clusters may vary significantly leading to clusters showing, however, similar signatures in calorimetry and in terms of hardening capability.

## Acknowledgements

The authors would like to thank Dr Christophe Sigli for providing the alloys and fruitful discussions. The authors acknowledge the financial support from Constellium C-Tec. The staff of Institut Laue Langevin (ILL) at beamline D11 is thanked for helping with the SANS experiments (doi:10.5291/ILL-DATA.1-01-142).

| Alloy | Cu at. % (wt. %) | Li at. % (wt. %) | Mg at. % (wt. %) | Al | Mg/Cu |
|---|---|---|---|---|---|
| Al-Cu-Mg | 1.1 (2.5) | --N/A-- | 1.7 (1.5) | Bal. | 1.55 (0.6) |
| Al-Cu-Li | 1.5 (3.5) | 3.5 (0.9) | --N/A-- | Bal. | --N/A-- |
| Al-Cu-Li-Mg | 1.5 (3.5) | 3.5 (0.9) | 0.45 (0.4) | Bal. | 0.3 (0.11) |

*Table 1: Alloy compositions*

| | Cu | Mg | Li |
|---|---|---|---|
| SAXS | 1.23 | -0.08 | -0.77 |
| SANS | 1.23 | 0.55 | -1.55 |

*Table 2: SAS elemental sensitivity in aluminium alloys as defined by $(f_i - f_{Al})/f_{Al}$, with $f_i$ and $f_{Al}$ the scattering factors of element i and Al, respectively*

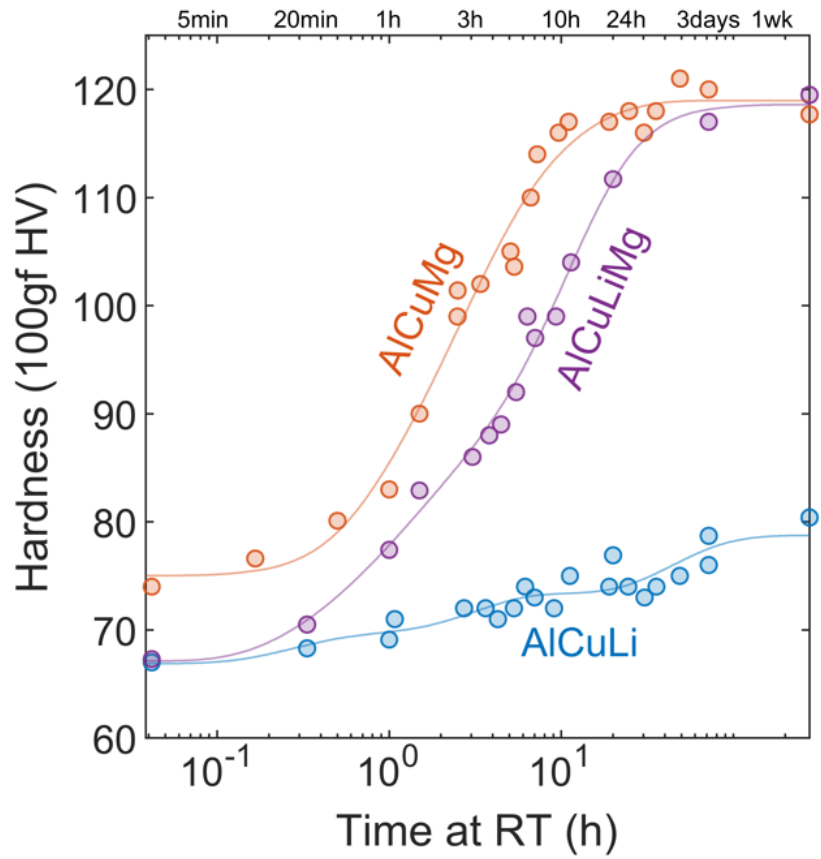

*Figure 1: Natural age hardening curves for Al-Cu-Mg, Al-Cu-Li and Al-Cu-Li-Mg alloys.*

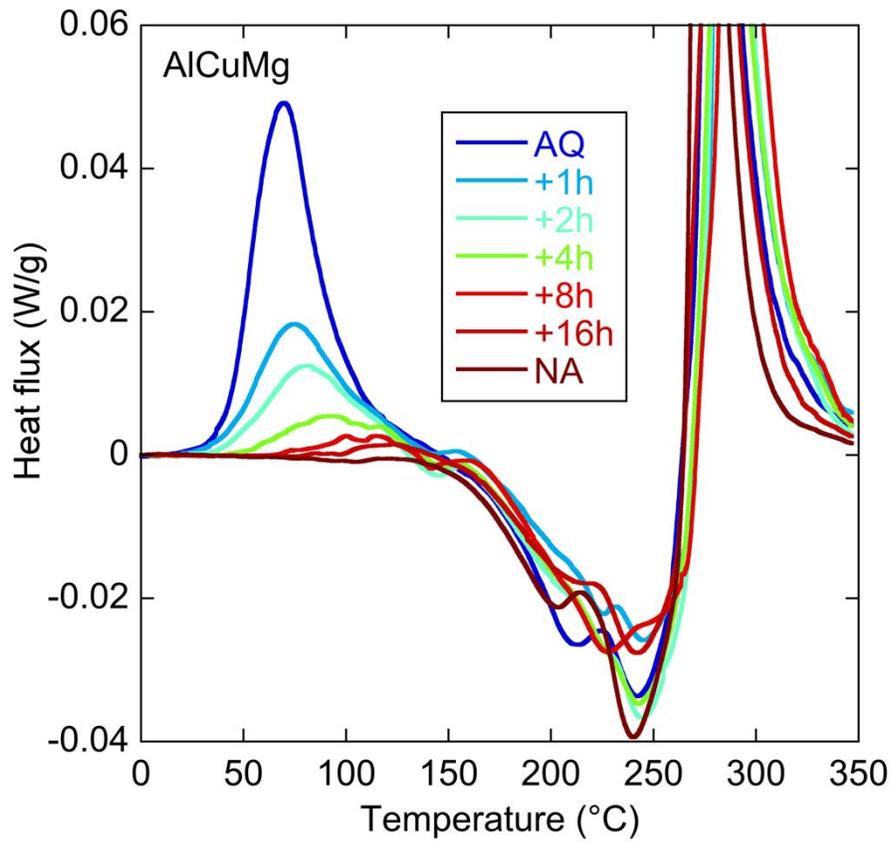

*Figure 2: DSC heat flux profiles for Al-Cu-Mg alloy carried out after various ageing times at room temperature with a heating rate of 10 °C/min. The AQ condition represents a case where all clusters are forming and dissolving in the DSC. The NA condition represents the case where all clusters have formed at room temperature and are only dissolving in the DSC.*

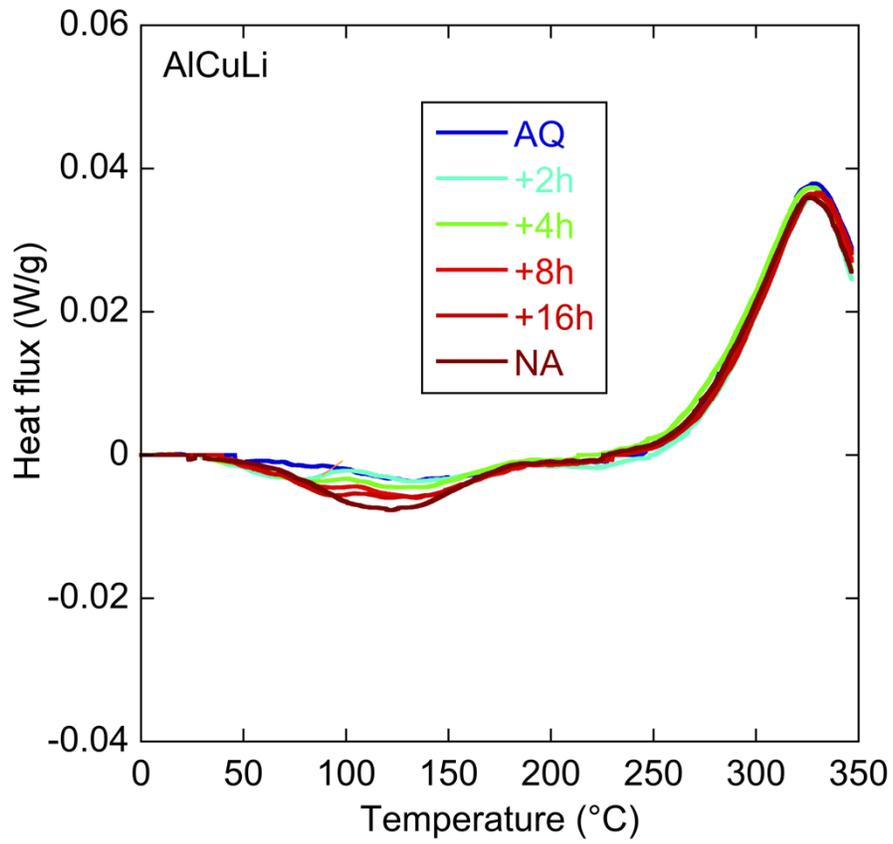

*Figure 3: DSC heat flux profiles for Al-Cu-Li alloy carried out after various ageing times at room temperature with a heating rate of 10 °C/min.*

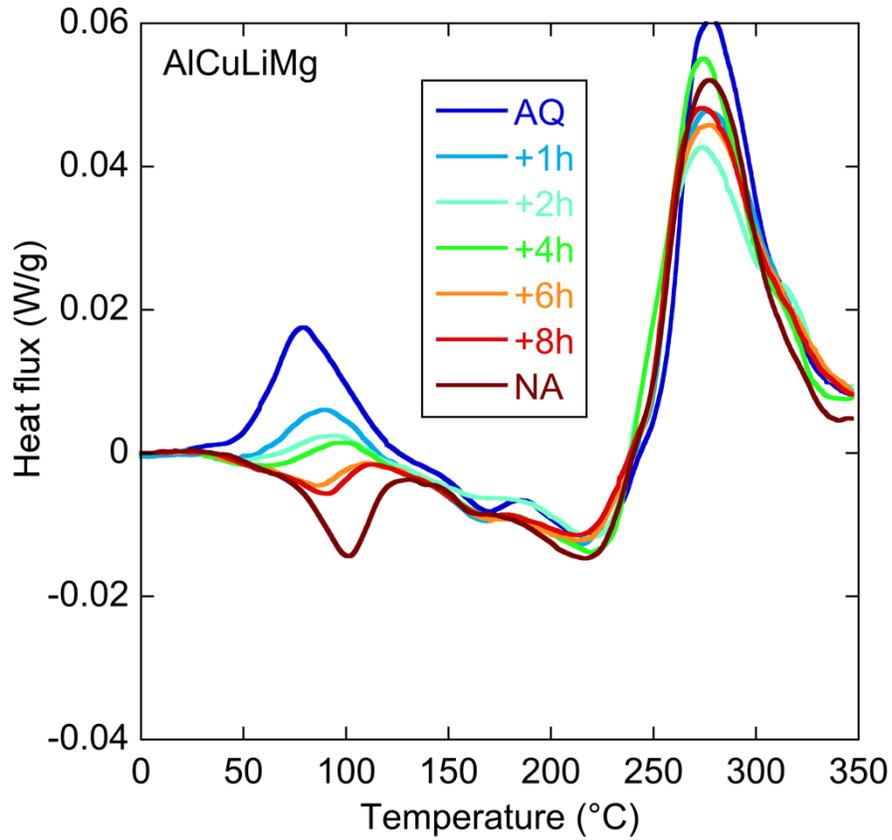

*Figure 4: DSC heat flux profiles for Al-Cu-Li-Mg alloy carried out after various ageing times at room temperature with a heating rate of 10 °C/min. The AQ condition represents a case where clusters are forming and dissolving in the DSC. The NA condition represent the case where clusters have formed at room temperature and are only dissolving in the DSC. The cluster reactions are significantly influenced by minor additions of Mg in comparison to Figure .*

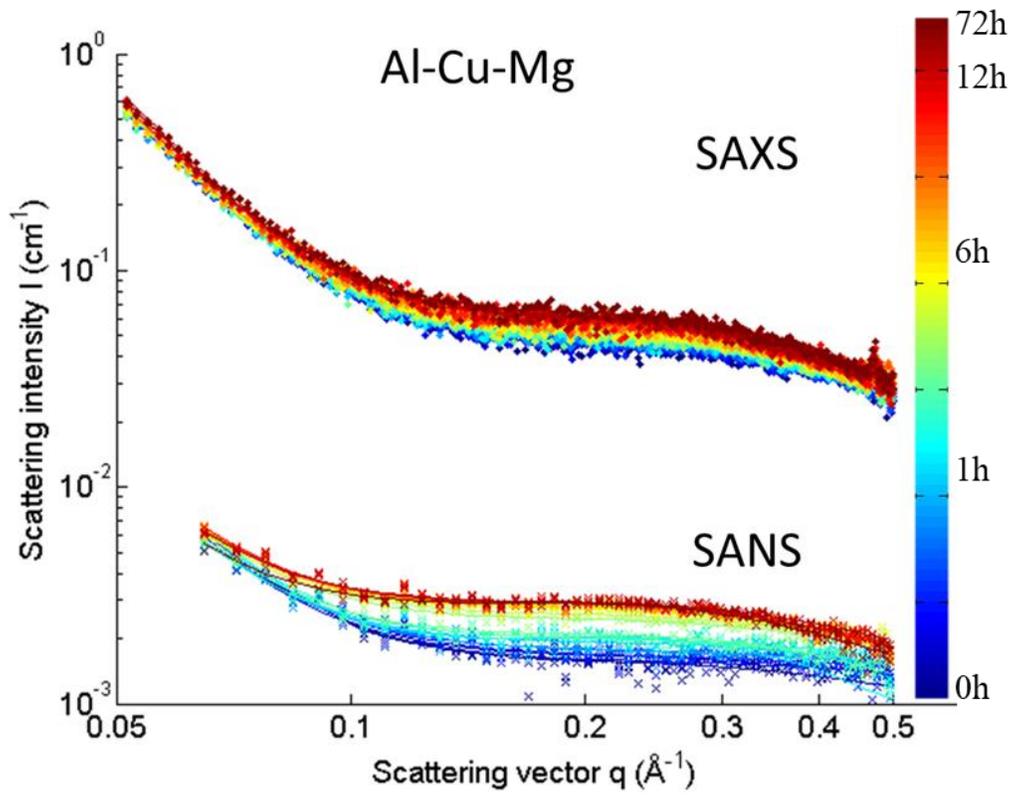

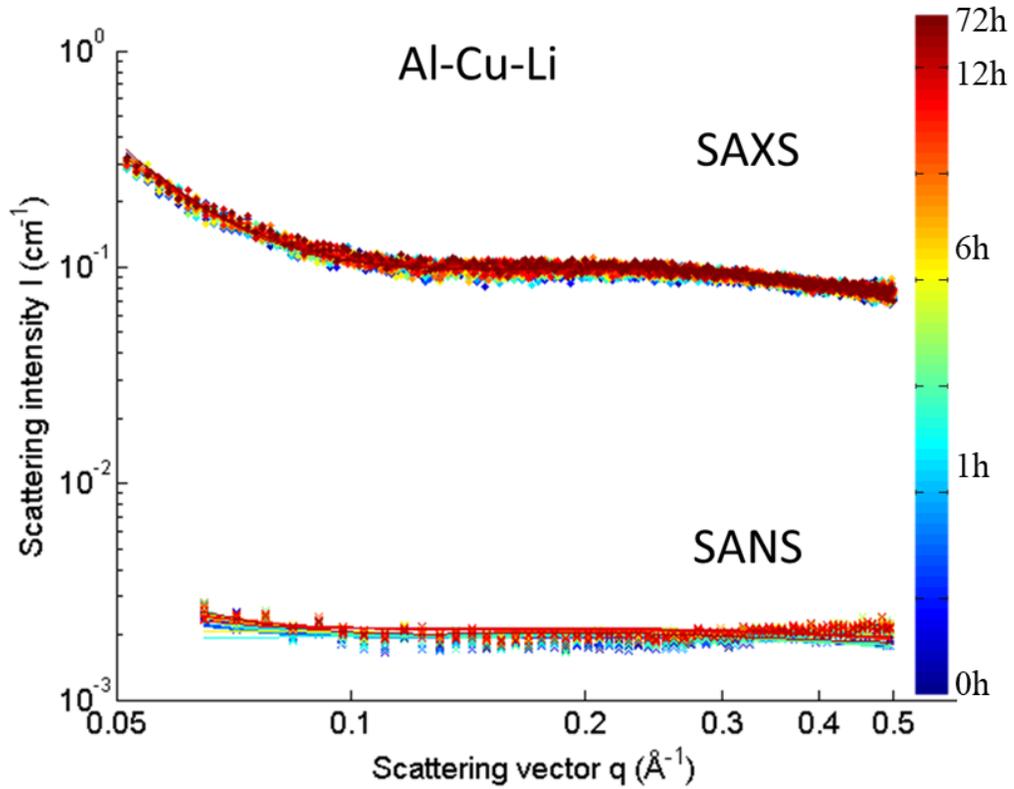

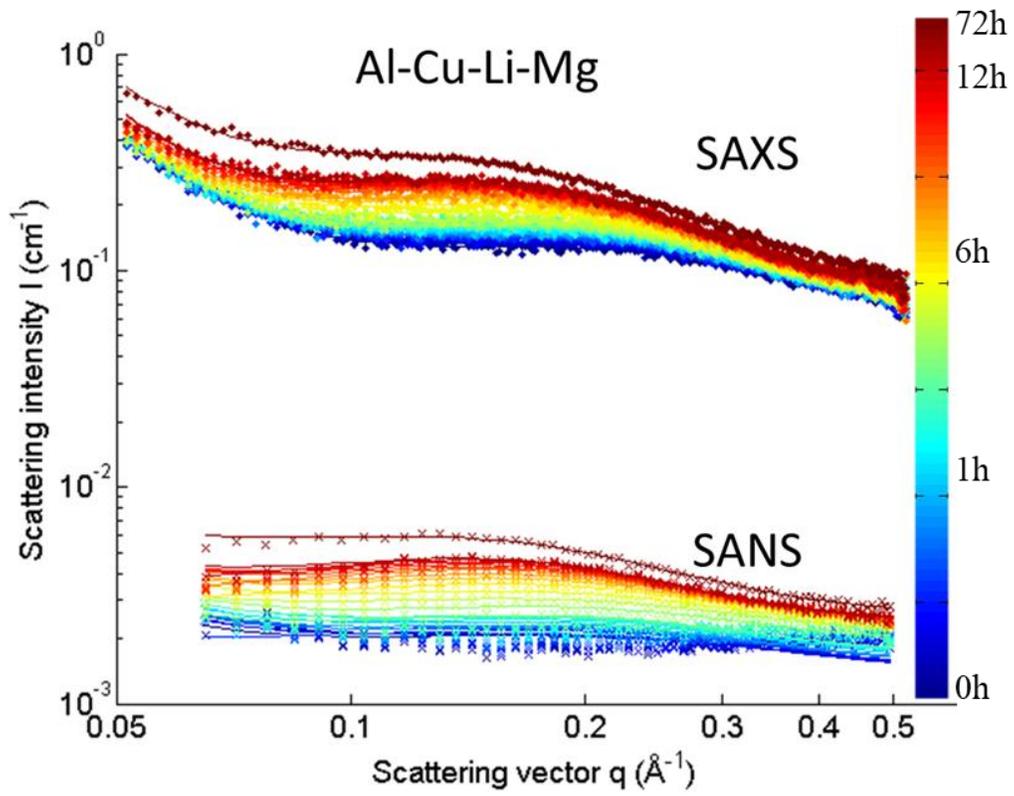

*Figure 5: Radially averaged scattering data evolution for a) Al-Cu-Li, b) Al-Cu-Mg and c) Al-Cu-Li-Mg alloys during in-situ SAXS (•) and SANS (x) experiments. Solid lines are model fits for the data.*

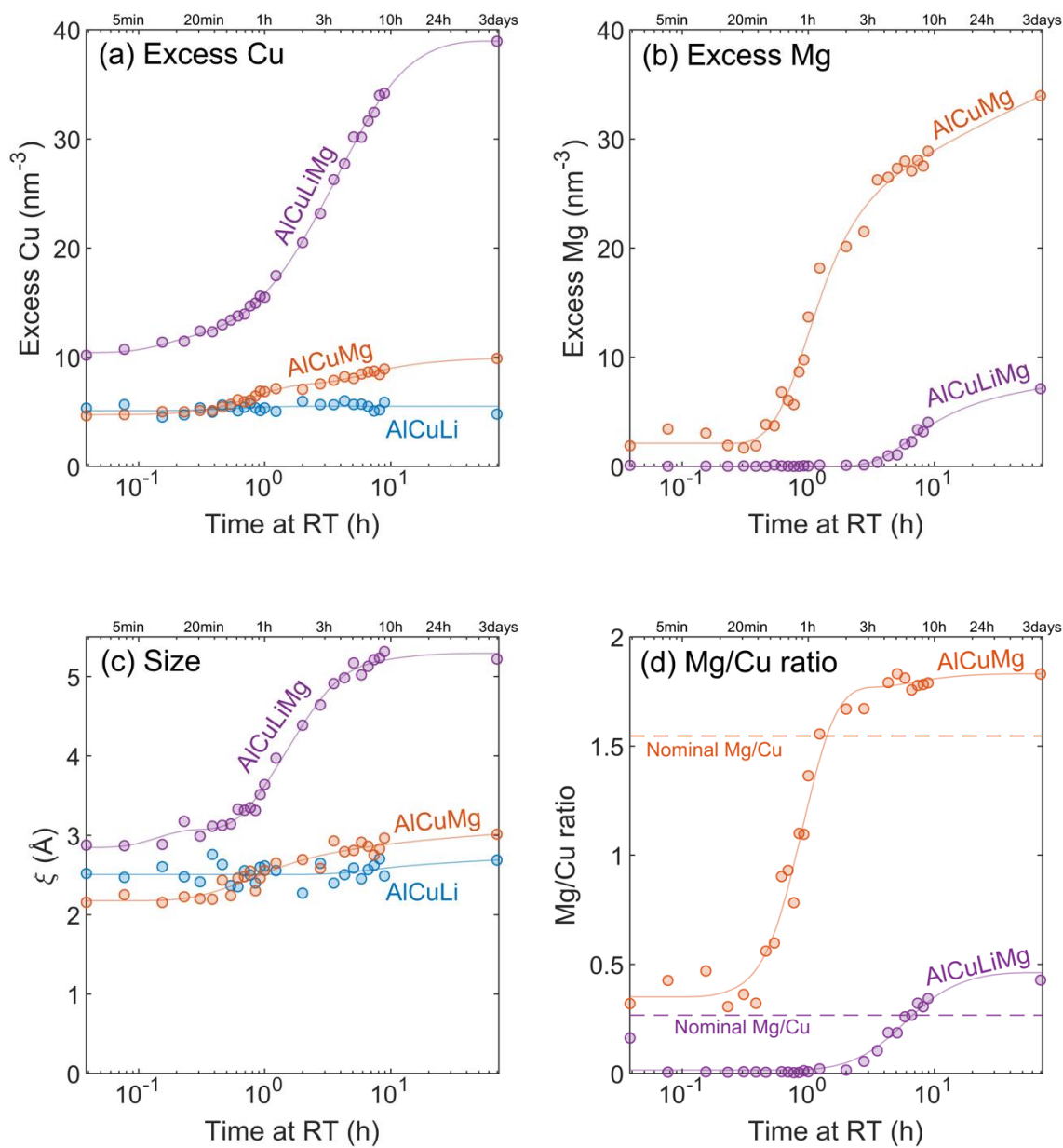

Figure 6 Evolution of the clusters characteristics during natural ageing for the three alloys: (a) Mean square number density of excess Cu atoms (b) Mean square number density of excess Mg atoms (c) Cluster size as defined by the correlation length (d) Mg/Cu ratio in the clusters; the dashed lines correspond to the nominal ratio in the two alloys containing Mg.